\documentclass[12pt]{article}

\usepackage[utf8]{inputenc}
\usepackage{latexsym, graphicx} 
\usepackage{amsmath,amsfonts,amssymb}
\usepackage{slashed}
\usepackage{cancel}
\usepackage{mathrsfs}
\usepackage{tensor}
\usepackage{xcolor}
\usepackage{cite}
\usepackage[colorlinks=true,linkcolor=blue,citecolor=blue,urlcolor=blue,filecolor=black]{hyperref}

\renewenvironment{thebibliography}[1]{
	\begin{oldthebibliography}{#1}
		\setlength{\itemsep}{2pt}
		\setlength{\parskip}{2pt}
	}
	{
	\end{oldthebibliography}
}

\numberwithin{equation}{section} 

\newcommand*{\scri}{\ensuremath{\mathscr{I}}}
\newcommand*{\lied}{\mathop{}\!\mathcal{L}}
\newcommand*{\dd}{\mathop{}\!d}

\begin{document}
	
	\begin{titlepage}
		\thispagestyle{empty}
		
		\begin{flushright}
		\end{flushright}
		
		\vskip1cm
		
		\begin{center}  
			{\Large\textbf{Carrollian conformal fields and flat holography}}
			
			\vskip1cm
			
			\centerline{Kevin Nguyen$^\dagger$ and Peter West$^\ddag$}
			
			\vskip1cm
			
			{\it{$^{\dagger\ddag}$Department of Mathematics, King's College London,\\
					The Strand, London WC2R 2LS, UK}}\\
			\vskip 0.2cm
			{\it{$^\ddag$Mathematical Institute, University of Oxford,\\
					Woodstock Road, Oxford, OX2 6GG, UK}}\\
			\vskip 0.5cm
			{kevin.nguyen@kcl.ac.uk, peter.west540@gmail.com}
			
		\end{center}
		
		\vskip1cm
		
		\begin{abstract} 
			The null conformal boundary $\scri$ of Minkowski spacetime $\mathbb{M}$ plays a special role in scattering theory, as it is the locus where massless particle states are most naturally defined. We construct quantum fields on $\scri$ which create these massless states from the vacuum and transform covariantly under Poincar\'e symmetries. Since the latter symmetries act as Carrollian conformal isometries of $\scri$, these quantum fields are Carrollian conformal fields. This group theoretic construction is intrinsic to $\scri$ by contrast to existing treatments in the literature. However we also show that the standard relativistic massless quantum fields in $\mathbb{M}$, when pulled back to $\scri$, provide a realisation of these Carrollian conformal fields. This correspondence between bulk and boundary fields should constitute a basic entry in the dictionary of flat holography. Finally we show that $\scri$ provides a natural parametrisation of the massless particles as described by irreducible representations of the Poincar\'e group, and that in an appropriate conjugate basis they indeed transform like Carrollian conformal fields. 
		\end{abstract}
		
	\end{titlepage}
	
	{\hypersetup{linkcolor=black}
		\tableofcontents 
	}
	
	\section{Introduction}
	\label{sec:intro}
	Central to the edifice of quantum field theory is the construction of states furnishing unitary representations of symmetry groups, together with the construction of local and covariant quantum fields creating these states out of the vacuum. For the important case of the Poincar\'e group $ISO(1,3)$, the unitary irreducible representations (UIR) were constructed by Wigner through the method of induced representations \cite{Wigner:1939cj}. A systematic construction of covariant quantum fields was provided only later by Weinberg \cite{Weinberg:1964cn,Weinberg:1964ev,Weinberg:1964kqu}, while Lagrangian field equations were given in \cite{Singh:1974qz,Singh:1974rc,Fronsdal:1978rb,Fang:1978wz}. However it is important to emphasise that a choice is typically being made when constructing these fields, namely that they should be functions over Minkowski spacetime $\mathbb{M}$, while this is not the only available option. In this paper we wish to consider an alternative to this standard procedure, namely the construction of local and covariant quantum fields as functions defined over null infinity~$\scri=\mathbb{R} \times \mathbb{R}^2$ which is the null conformal boundary of $\mathbb{M}$. Our discussion will in fact apply to arbitrary spacetime dimension and corresponding symmetry group $ISO(1,d)$. 
	
	The main motivation for the present work comes from the program of \textit{flat/celestial holography}, which aims at a better understanding of asymptotically flat quantum gravity by exploiting the vast amount of asymptotic symmetries available, and by use of conformal methods. For an account of this subject we refer the reader to the reviews  \cite{Strominger:2017zoo,Pasterski:2021raf,McLoughlin:2022ljp} and references therein. One basic but important aspect of flat holography is to provide a dictionary between standard relativistic quantum fields in $\mathbb{M}$ and conformal quantum fields at $\scri$. In that respect two approaches have been pursued, either by further reducing $\scri$ along its null direction down to the two-dimensional Riemann sphere, or by keeping this null direction manifest. Most of the literature has followed the first approach, with the result that one associates a continuum of two-dimensional $SL(2,\mathbb{C})$ primary fields to a given relativistic bulk field \cite{deBoer:2003vf,Cheung:2016iub,Pasterski:2016qvg,Pasterski:2017kqt,Pasterski:2017ylz,Lam:2017ofc,Banerjee:2018gce,Fotopoulos:2019vac,Law:2020tsg,Iacobacci:2020por,Pasterski:2020pdk,Narayanan:2020amh,Pano:2021ewd}. The second approach has the definite advantage that full $ISO(1,3)$ covariance is kept manifest, with the action of the Poincar\'e group realised as conformal isometries of $\scri$. In that case a single Carrollian\footnote{The term \textit{Carrollian}, by contrast to the term \textit{relativistic}, refers to the degenerate nature of $\scri$ endowed with spacetime signature $(0,+,+)$. Investigations of Carrollian conformal field theories include \cite{Bagchi:2016bcd,Bagchi:2019xfx,Bagchi:2019clu,Chen:2021xkw,Bagchi:2023fbj,Salzer:2023jqv}.} conformal field at $\scri$ is found to correspond to a given relativistic quantum field in $\mathbb{M}$ \cite{Bagchi:2010zz,Donnay:2022aba,Donnay:2022wvx,Bagchi:2022emh,Saha:2023hsl,Bagchi:2023fbj}. However we believe that a systematic discussion of the correspondence between relativistic bulk fields, Carrollian conformal fields and particle states is to some extent incomplete, and we intend to close this gap here.\footnote{However we restrict the discussion to massless fields. Massive fields require an entirely different description as they cannot be realised as Carrollian conformal fields over $\scri$ as \eqref{C2=0} clearly demonstrates. The underlying physical reason is that massive states propagate to timelike infinity $i^+$ rather than $\scri$. It is therefore more natural to try and construct massive Carrollian fields living on (a blow-up) of $i^+$. We leave this for future investigation.} Nonetheless our analysis is complementary and has been inspired by previous works \cite{Bagchi:2010zz,Bagchi:2016bcd,Donnay:2022aba,Donnay:2022wvx,Bagchi:2022emh}. We claim that the holographic dictionary between relativistic fields in $\mathbb{M}$ and Carrollian conformal fields on $\scri$ can be deduced solely from group theoretic considerations. All that is required is to understand how to turn Wigner's UIRs into local covariant quantum fields defined over either $\mathbb{M}$ or $\scri$. The holographic correspondence then relates the two types of fields (relativistic and Carrollian) that are associated to a given UIR. 
	
	The paper is structured as follows. In section~\ref{sec:helicity} we recall the form of the covariant quantum fields in $\mathbb{M}^{d+1}$ which create the massless representations of $ISO(1,d)$, together with the free field equations that they satisfy. In section~\ref{sec:boundary} we independently construct the Carrollian conformal fields on $\scri=\mathbb{R} \times \mathbb{R}^{d-1}$ that can be associated with the very same massless representations, exploiting the fact that $ISO(1,d)$ is the group of conformal isometries of $\scri$. Just like in the standard case of relativistic conformal fields \cite{Mack:1969rr}, this is done through the method of induced representations, starting from a finite-component representation of the stability subgroup of the origin of $\scri$. We further show that Carrollian fields at $\scri$ can only possibly create massless states, which is how group theory tells us that massive fields are supported at future timelike infinity $i^+$ rather than at null infinity. In section~\ref{sec:bulk} we explicitly connect the two independent constructions by showing that the Carrollian conformal fields arise as asymptotic limits of the standard relativistic massless fields. To be more specific, we adopt retarded coordinates $(r,u,x^i)$ such that null infinity $\scri$ corresponds to the limit $r \to \infty$. In that limit we show that the independent gauge-invariant components of the relativistic massless fields precisely behave as Carrollian conformal fields under $ISO(1,d)$ transformations. This provides an explicit dictionary between relativistic massless fields in $\mathbb{M}^{d+1}$ and Carrollian conformal fields at $\scri$. In section~\ref{sec:particle states} we provide a more direct relation between the massless particle states and the Carrollian conformal fields. First the irreducible massless representations of the Poincar\'e group are constructed  in momentum space by boosting the rest frame states using the generators from outside the little group as originally set out by Wigner. Taking their Fourier transform to Minkowski spacetime and then the $r\to \infty $ limit the corresponding states on the boundary $\scri$ are found.  It is shown that the parametrisation of the states is very closely related to the coordinates $(u,x^i)$ on $\scri$. Indeed the coordinates $x^i$ are precisely the same as those in the parametrisation of the boost while the remaining coordinate $u$ is a conjugate coordinate in the sense of Fourier transform. Indeed  $\scri$ encodes in an unconstrained way the kinematics and the states of a massless particle, essentially as they  appear in the irreducible representation of the Poincar\'e group which characterises it.  The physical states that emerge on $\scri$ transform in a way reminiscent to the Carrollian conformal fields of section~\ref{sec:boundary}. Embedding the physical states in a larger representation of the type used to display the Poincar\'e transformations in a manifest manner we find states on $\scri$ 
	that now transform exactly like the Carrollian conformal fields constructed in section~\ref{sec:boundary}.
	
	\section{Relativistic massless fields}
	\label{sec:helicity}
	Fundamental to the standard scattering theory in $(d+1)$-dimensional Minkowski spacetime $\mathbb{M}^{d+1}$ is the assumption that asymptotic states belong to the tensor product of unitary irreducible representations (UIR) of the isometry group $ISO(1,d)$, and more specifically those induced by finite-dimensional representations of the corresponding (short) little groups \cite{Wigner:1939cj,Weinberg:1995mt}. In the case of massless states the \textit{helicity representations} are those which carry finite-dimensional UIRs of the short little group $SO(d-1)$. Single-particle states $|p,\sigma\rangle$ belonging to a given helicity representation are labeled by a null momentum $p^\mu$ and a discrete label $\sigma$ running over the internal spin degrees of freedom. Furthermore they are built out of the vacuum by the corresponding creation operators,
	\begin{equation}
	|p,\sigma\rangle \equiv a^\dagger_{p,\sigma}\, |0\rangle\,.
	\end{equation}
	
	In the present work we will restrict our attention to \textit{totally symmetric tensor} helicity representations. In four spacetime dimensions ($d=3$) totally symmetric representations are the only available ones, while in higher dimensions there is also the possibility of mixed symmetry \cite{Labastida:1986ft,Labastida:1986gy,Labastida:1987kw,Bekaert:2002dt,Bekaert:2006ix,Bekaert:2006py}.
	
	To a given helicity representation one can associate a local quantum field which creates single-particle states out of the vacuum, and transforms covariantly albeit non-unitarily. This local quantum field takes the generic form,
	\begin{equation}
	\label{intertwining}
	\phi_{\mu_1...\mu_s}(x)=\sum_\sigma \int d^dp\, \phi^{p,\sigma}_{\mu_1...\mu_s}(x)\, a^\dagger_{p,\sigma}\,, 
	\end{equation}
	with $d^dp$ a Lorentz-invariant measure on the lightcone $p^2=0$. The \textit{intertwiners} $\phi^{p,\sigma}_{\mu_1...\mu_s}(x)$ are determined precisely such that the unitary transformation of the creation operators $a^\dagger_{p,\sigma}$ is converted into covariant transformation of the local quantum field $\phi_{\mu_1...\mu_s}(x)$ \cite{Weinberg:1995mt}. It turns out that these intertwiners are just the (gauge-invariant) positive-frequency solutions to a set of covariant field equations. Here we recall the form of these equations and refer the reader to \cite{Bekaert:2006py,Rahman:2015pzl,Ponomarev:2022vjb} for a complete review. The quantum field associated with a spin-$s$ helicity representation is a totally symmetric tensor $\phi_{\mu_1...\mu_s}$ satisfying the covariant field equations
	\begin{align}
	\label{bulk eom phi}
	\square \phi_{\mu_1...\mu_s}=0\,, \qquad \nabla^{\nu} \phi_{\nu\mu_2...\mu_s}=0\,, \qquad  \phi^{\nu}{}_{\nu\mu_3...\mu_s}=0\,,
	\end{align}
	subject to the gauge redundancy
	\begin{align}
	\label{gauge symmetry}
	\delta \phi_{\mu_1...\mu_s}=\nabla_{(\mu_1} \varepsilon_{\mu_2...\mu_s)}\,,
	\end{align}
	where the totally symmetric gauge parameter $\varepsilon_{\mu_1...\mu_{s-1}}$ itself satisfies
	\begin{align}
	\label{bulk eom xi}
	\square \varepsilon_{\mu_1...\mu_{s-1}}=0\,, \qquad \nabla^{\nu} \varepsilon_{\nu\mu_2...\mu_{s-1}}=0\,, \qquad  \varepsilon^{\nu}{}_{\nu\mu_3...\mu_{s-1}}=0\,.
	\end{align}
	The field equations \eqref{bulk eom phi} constitute a partially gauge-fixed version of Fronsdal equations \cite{Fronsdal:1978rb}. While the wave equation is the Casimir equation $\mathcal{C}_2\equiv P^\mu P_\mu=0$ characterising massless states, the remaining equations can be used to show that the gauge-invariant tensor components precisely reduce to that of a traceless symmetric tensor $\phi_{i_1...i_s}$ furnishing a spin-$s$ representation of the short little group $SO(d-1)$ \cite{deWit:1979sib}. 
	
	The notion of locality and covariance discussed above is that of local tensor fields in $\mathbb{M}^{d+1}$. We will be interested in a distinct notion of covariance with respect to a null plane $\scri=\mathbb{R} \times \mathbb{R}^{d-1}$ of codimension one, where the map between those two descriptions is exactly what should constitute the basics of the flat holographic dictionary.  
	
	\section{Carrollian conformal fields}
	\label{sec:boundary}
	The Poincar\'e group $ISO(1,d)$ naturally acts as the group conformal isometries\footnote{The full group of conformal isometries of $\scri$ can be larger than $ISO(1,d)$, depending on the dimension and on the precise geometric structure which one tries to preserve \cite{Duval:2014lpa,Duval:2014uva}. For the particularly relevant case of $d=3$, the group of conformal isometries of the pair $(q_{\alpha\beta},n^\alpha)$ is the extended BMS group that contains supertranslations and superrotations \cite{Bondi:1962px,Sachs:1962wk,Barnich:2009se}. We refer the reader to \cite{Nguyen:2022zgs} for a concise review which includes the connection to four-dimensional gravity. In the present paper we restrict our attention to $ISO(1,d)$ since it underlies the construction of the standard scattering states.} of a generic null plane $\scri=\mathbb{R}\times \mathbb{R}^{d-1}$ equipped with the canonical metric 
	\begin{equation} 
	ds^2_{\scri}=q_{\alpha\beta} \dd x^\alpha \dd x^\beta=0\, \dd u^2 +\delta_{ij} \dd x^i \dd x^j\,.
	\end{equation}
	This metric is degenerate, i.e., there is a vector $n^\alpha$ such that
	\begin{equation}
	n^\alpha q_{\alpha\beta}=0\,.
	\end{equation}
	In the coordinate system $x^\alpha=(u,x^i)$, this vector is simply given by $n^\alpha=(1,0^i)$. 
	
	Perhaps the easiest way to see that $ISO(1,d)$ is the conformal group associated with $\scri$ is to obtain it as an In\"on\"u-Wigner contraction of the conformal group $SO(2,d)$ of $\mathbb{M}^d$. Historically this is exactly how L\'evy-Leblond introduced the notion of Carrollian isometries \cite{Levy1965}. The relevant contraction corresponds to the ultrarelativistic/Carrollian limit $c \to 0$ upon which
	\begin{equation}
	\mathbb{M}^d \to \scri\,, \qquad SO(2,d) \to ISO(1,d)\,.
	\end{equation}
	This contraction is explicitly performed in the appendix and allows to derive the subsequent formulae, some of which can also be found in \cite{Bagchi:2016bcd,Bagchi:2022emh}.
	
	Infinitesimally the Carrollian conformal isometries under consideration take the form $x'^{\alpha}=x^\alpha+\zeta^\alpha$ in terms of the vector field
	\begin{equation}
	\label{zetas}
	\begin{split}
	\zeta^u&=a+b_i x^i+k x^2+(\lambda-2 k_i x^i)u\,,\\
	\zeta^i&=a^i+\omega^i{}_j\, x^j+\lambda x^i+k^i x^2-2k_j x^j x^i\,.
	\end{split}
	\end{equation}
	Alternatively this can be written
	\begin{align}
	x'^{\alpha}&=(1+i a H+i a^i P_i+\frac{i}{2} \omega^{ij} J_{ij}+i b^i B_i+i \lambda D+i k K+i k^i K_i)\, x^{\alpha}\,,
	\end{align}
	in terms of the differential operators
	\begin{align}
	\nonumber
	P_i&=-i \partial_i\,, & J_{ij}&=i(x_i \partial_j-x_j \partial_i )\,,\\
	\label{Carroll generators}
	D&=-i(u\partial_u+x^i\partial_i )\,, & K_i&=-i(x^2 \partial_i-2x_i(u\partial_u+x^j\partial_j))\,, & K&=-i x^2 \partial_u\,,\\
	\nonumber
	H&=-i \partial_u\,, & B_i&=-i x_i \partial_u\,, 
	\end{align}
	that satisfy the algebra
	\begin{align}
	\nonumber
	\left[J_{ij}\,,J_{mn}\right]&=-i\left(\delta_{im} J_{jn}+\delta_{jn} J_{im}-\delta_{in} J_{jm}-\delta_{jm} J_{in} \right)\,, & \left[D\,, P_i \right]&=i P_i\,,\\
	\nonumber
	\left[J_{ij}\,,P_k \right]&=-i \left(\delta_{ik} P_j-\delta_{jk} P_i \right)\,, & \left[D\,, H \right]&=i H\,,\\
	\nonumber
	\left[J_{ij}\,,K_k \right]&=-i \left(\delta_{ik} K_j-\delta_{jk} K_i \right)\,, & \left[D\,, K_i \right]&=-i K_i\,,\\
	\label{conformal Carroll algebra}
	\left[J_{ij}\,,B_k \right]&=-i\left(\delta_{ik}B_j-\delta_{jk} B_i\right)\,, & \left[D\,, K \right]&=-i K\,,\\
	\nonumber
	\left[B_i\,, P_j \right]&=i \delta_{ij} H\,, & \left[H\,, K_i\right]&=2i B_i\,,\\
	\nonumber
	\left[B_i\,,K_j \right]&=i \delta_{ij} K\,, & \left[K\,, P_i\right]&=2i B_i\,,\\
	\nonumber
	\left[K_i\,, P_j\right]&=-2i\left(\delta_{ij} D-J_{ij} \right)\,.
	\end{align}
	This algebra is isomorphic to $\mathfrak{iso}(1,d)$, which can be seen explicitly through the identification
	\begin{equation}
	\label{tilde J}
	\tilde J_{ij}=J_{ij}\,, \qquad \tilde J_{i0}=-\frac{1}{2}\left(P_i+K_i \right)\,, \qquad \tilde J_{id}= \frac{1}{2}\left(P_i-K_i \right)\,, \qquad \tilde J_{0d}=-D\,,
	\end{equation}
	and
	\begin{equation}
	\label{tilde P}
	\tilde P_0=\frac{1}{\sqrt{2}}(H+K)\,, \qquad \tilde P_i=-\sqrt{2}\, B_i\,, \qquad \tilde P_d=\frac{1}{\sqrt{2}}(K-H)\,,
	\end{equation}
	such that $\{\tilde J_{\mu\nu}\,, \tilde P_\mu \}$ indeed satisfy the $\mathfrak{iso}(1,d)$ algebra in its standard form. Crucially this representation of the Poincar\'e algebra is such that the quadratic Casimir operator vanishes identically,
	\begin{equation}
	\label{C2=0}
	\mathcal{C}_2=\tilde P^\mu \tilde P_\mu=-\tilde P_0^2+\tilde P^i \tilde P_i+\tilde P_d^2=-(HK+KH)+2B^i B_i=2x^2 \partial_u^2-2x^2 \partial_u^2=0\,.
	\end{equation}
	This is in stark contrast with the standard representation $\mathcal{C}_2=\partial^\mu \partial_\mu$ associated with Minkowski space $\mathbb{M}^{d+1}$. This shows that fields at $\scri$ can only possibly carry massless representations of $ISO(1,d)$.
	
	We now turn to the construction of fields defined locally on $\scri$ and transforming covariantly under $ISO(1,d)$. Just like in the standard case of relativistic conformal fields \cite{Mack:1969rr}, we first look for finite-dimensional irreducible representations of the stability subgroup $H$ of the origin $x^\alpha=0$. Looking at \eqref{Carroll generators} we see that the latter is generated by the algebra
	\begin{equation}
	\mathfrak{h}=\{J_{ij}, B_i,K_i,K,D \}\,.
	\end{equation}
	First we note that $J_{ij}$ are $SO(d-1)$ generators, which naturally leads us to consider a symmetric and traceless spin-$s$ tensor field $\phi(0)\equiv \phi_{i_1...i_s}(0)$ transforming like
	\begin{equation}
	\label{J B actions}
	\left[J_{ij}, \phi(0)\right]=\Sigma_{ij}\, \phi(0)\,, 
	\end{equation}
	where $\Sigma_{ij}$ is the standard $SO(d-1)$ irreducible spin-$s$ hermitian representation. This exactly coincides with the UIR of the short little group $SO(d-1)$ from which the spin-$s$ helicity representation is induced. Since $B_i$ and $K_i$ transform like $SO(d-1)$ vectors, finite dimensionality of $\phi(0)$ requires them to act trivially,
	\begin{equation}
	\label{Bi Ki action}
	\left[B_i,\phi(0)\right]=\left[K_i,\phi(0)\right]=0\,.
	\end{equation}
	Consistency with the algebra \eqref{conformal Carroll algebra} then requires the generator $K$ to act trivially too,
	\begin{equation}
	\label{K action}
	\left[B_i,K_j\right]=i\delta_{ij} K \quad \Rightarrow \quad \left[K,\phi(0)\right]=0\,.
	\end{equation}
	On the other hand, since it commutes with the spin generators the action of the dilation operator can be diagonalised,
	\begin{equation}
	\label{action D}
	\left[D, \phi(0)\right]=i\Delta\, \phi(0), \qquad \Delta \in \mathbb{R}\,.
	\end{equation}
	Note that $K_\alpha=(K,K_i)$ and $P_\alpha=(H,P_i)$ act as lowering and raising operator for the conformal dimension, respectively,
	\begin{equation}
	\begin{split}
	\left[D,\left[K_\alpha, \phi(0)\right] \right]&=i(\Delta-1)\left[K_\alpha, \phi(0) \right]\,,\\
	\left[D,\left[P_\alpha, \phi(0)\right] \right]&=i(\Delta+1)\left[P_\alpha, \phi(0) \right]\,,
	\end{split}
	\end{equation}
	and the triviality of $K_\alpha$ imposed in \eqref{Bi Ki action}-\eqref{K action} amounts to the \textit{primary field condition}
	\begin{equation}
	\label{K=0}
	\left[K_\alpha, \phi(0)\right]=0\,.
	\end{equation}
	As we argued this is a logical consequence of the finite dimensionality of $\phi(0)$. 
	
	The dependence on the coordinates $x^\alpha=(u,x^i)$ is then obtained by use of the translation operators $P_\alpha=(H,P_i)$,
	\begin{equation}
	\phi(x)\equiv U(x)\, \phi(0)\, U(x)^{-1}\,, \qquad U(x)\equiv e^{-i x^\alpha P_\alpha}= e^{-i(u H+x^i P_i)}\,.
	\end{equation}
	To work out the action of an arbitrary generator $X \in \mathfrak{iso}(1,d)$ on the field $\phi(x)$, we make use of 
	\begin{equation}
	\left[X, \phi(x)\right]=U(x) \left[X', \phi(0)\right] U(x)^{-1}\,,
	\end{equation}
	where
	\begin{equation}
	\label{X prime}
	X'=U(x)^{-1}XU(x)=\sum_{n=0}^\infty \frac{i^n}{n!} x^{\alpha_1}\,...\,x^{\alpha_n} \left[P_{\alpha_1},\left[\,...\,\left[P_{\alpha_n},X\right]\right] \right]\,.
	\end{equation}
	Explicitly using the form \eqref{conformal Carroll algebra} of the $\mathfrak{iso}(1,d)$ algebra, this series truncates at order $n=2$ at most, and yields the infinitesimal action
	\begin{equation}
	\label{Carrollian induced rep}
	\begin{split}
	\left[H, \phi(x)\right]&=-i \partial_u \phi(x)\,,\\
	\left[P_i, \phi(x)\right]&=-i \partial_i \phi(x)\,,\\
	\left[J_{ij}, \phi(x)\right]&=-i\left(i\Sigma_{ij}-x_i \partial_j+x_j\partial_i \right)\phi(x)\,,\\
	\left[D, \phi(x)\right]&=-i\left(-\Delta+u\partial_u+x^i\partial_i \right)\phi(x)\,,\\
	\left[K, \phi(x)\right]&=-i x^2\partial_u \phi(x)\,,\\
	\left[K_i, \phi(x)\right]&=-i\left(2x_i \Delta+2ix^j \Sigma_{ij}-2ux_i \partial_u-2x_i x^j \partial_j+x^2 \partial_i \right)\phi(x)\,,\\
	\left[B_i, \phi(x)\right]&=-ix_i\partial_u \phi(x)\,.
	\end{split}
	\end{equation}
	This can be compactly written in terms of the Lie derivative $\lied_\zeta$,
	\begin{equation}
	\label{Carrollian conformal transf}
	\begin{split}
	\delta \phi(x)&\equiv i[(a H+a^i P_i+\frac{1}{2}\omega^{ij}J_{ij}+b^i B_i+\lambda D+k K+k^i K_i)\,, \phi]\\
	&=\left(\zeta^\alpha \partial_\alpha-\frac{i}{2} \partial_{[i} \zeta_{j]}\, \Sigma^{ij}-\Delta \Omega\right)\phi= \left( \lied_\zeta -\Delta \Omega\right) \phi(x)\,,
	\end{split}
	\end{equation}
	where the scaling factor is given by
	\begin{equation}
	\label{Omega}
	\Omega=\frac{1}{d}\partial_\alpha \zeta^\alpha=\lambda-2k_i x^i\,.
	\end{equation}
	
	The Carrollian conformal field $\phi(x)$ transforms covariantly in an irreducible representation of $ISO(1,d)$. Up to this point their relation to the spin-$s$ helicity states is however unclear. To assess whether such covariant fields can carry the helicity states, we can fix a momentum frame and determine whether the independent field components transform in the UIR of the little group from which the full helicity representation is induced \cite{Bargmann:1948ck,Bekaert:2006py,Rahman:2015pzl,Ponomarev:2022vjb}. The point $x^\alpha=0$ corresponds to the momentum frame 
	\begin{equation}
	\label{momentum frame}
	\tilde P_\mu=\frac{1}{\sqrt{2}}(H,0,...,0,-H)\,,
	\end{equation}
	as follows from \eqref{J B actions} and \eqref{K=0}, and the massless character of the representation is again manifest. In that frame the little group which leaves \eqref{momentum frame} invariant is therefore generated by
	\begin{equation}
	\{ \tilde J_{ij}\,, \tilde J_{id}+\tilde J_{i0} \}=\{J_{ij}\,, K_i\}=\mathfrak{iso}(d-1)\,.
	\end{equation}
	Equations \eqref{J B actions} and \eqref{K=0} tell us that $\phi(0)$ carries the finite-dimensional spin-$s$ UIR of the little group, the very same UIR from which the spin-$s$ helicity representation is induced. Thus $\phi(x)$ is a good candidate quantum field to create the spin-$s$ helicity sates. The simple-minded argument given above does not immediately fix the conformal dimension $\Delta$ of the Carrollian field as the dilation operator $D$ does not belong to the little algebra $\mathfrak{iso}(d-1)$. Rather its action on massless particles is induced and therefore determined by the little group UIR. In section~\ref{sec:particle states} we directly build the Carrollian representations starting from the massless UIRs, which allows to determine the conformal dimension from the spin of the representation,
	\begin{equation}
	\label{Delta(s)}
	\Delta(s)=s-\frac{d-1}{2}\,.
	\end{equation}
	This will also be explicitly realised in section~\ref{sec:bulk} when pulling back relativistic bulk fields to future null infinity $\scri$.
	
	Finally we may wish to achieve general covariance, i.e., to embed $\phi_{i_1...i_s}$ into a Carrollian tensor $\phi_{\alpha_1...\alpha_s}$. This is achieved almost trivially by requiring the latter to be fully symmetric and by further imposing
	\begin{equation}
	\label{projection}
	n^{\alpha} \phi_{\alpha...\alpha_s}=0\,, \qquad q^{\alpha\beta} \phi_{\alpha\beta...\alpha_s}=0\,.
	\end{equation}
	Here $q^{\alpha \beta}$ is any symmetric tensor satisfying $q^{ij}=\delta^{ij}$, which is therefore defined only up to $q^{\alpha\beta} \mapsto q^{\alpha\beta}+l_1^\alpha n^\beta+n^\alpha l_2^\beta$ for any two vectors $l_{1,2}^\alpha$. However this ambiguity is inconsequential on account of the first condition in \eqref{projection}.
	Thus the only nonzero tensor components in $\phi_{\alpha_1...\alpha_s}$ are indeed the spatial components $\phi_{i_1...i_s}$. he transformation \eqref{Carrollian conformal transf} then takes the general covariant form
	\begin{align}
	\delta \phi_{\alpha_1...\alpha_s}&=\left(\zeta^\alpha \partial_\alpha-\frac{i}{2}\partial_{[\alpha} \zeta_{\beta]}\, \Sigma^{\alpha\beta}-\Delta \Omega\right)\phi_{\alpha_1...\alpha_s}=\left(\lied_\zeta-\Delta \Omega  \right)\phi_{\alpha_1...\alpha_s}\,,
	\end{align}
	where $\Sigma^{\alpha \beta}$ is any completion of $\Sigma^{ij}$.\footnote{The way the completion is achieved does not actually matter, given that the extra matrices $\Sigma^{ui}$ drop from the first line since $\zeta_u=0$ due to metric degeneracy and $\partial_u \zeta_i=0$, while the projection condition \eqref{projection} makes the extension of the matrices $\Sigma^{ij}$ in the $u$ directions irrelevant.} The last expression in terms of the Lie derivative $\lied_\zeta$ is obviously valid provided the projection condition \eqref{projection} holds.   
	
	This concludes the construction of finite-component conformal primary fields defined over a generic null plane $\scri$, and transforming in irreducible representations of the Carrollian conformal group $ISO(1,d)$. The above considerations were intrinsic to $\scri$ and made no reference whatsoever to a higher-dimensional Minkowskian geometry $\mathbb{M}^{d+1}$. Of course $\scri$ also arises as the conformal boundary of $\mathbb{M}^{d+1}$. In that context the goal of the next section will be to show that the pullback to $\scri$ of the standard bulk quantum fields provides realisations of the Carrollian conformal fields introduced above. It will also automatically provides the intertwining relation between these Carrollian quantum fields and the creation operators $a_{p,\sigma}^\dagger$ of the helicity representations.
	
	\section{The bulk-boundary dictionary}
	\label{sec:bulk}
	We now have a closer look at the standard massless quantum fields in $\mathbb{M}^{d+1}$ which were briefly described in section~\ref{sec:helicity}, and show that the Carrollian conformal fields constructed in section~\ref{sec:boundary} naturally arise as asymptotic values of these bulk fields. This is similar in spirit to \cite{Donnay:2022aba,Donnay:2022wvx} although the demonstration is different. To this end it is best to adopt retarded coordinates $(r,u,x^i)$, related to cartesian coordinates $X^\mu$ by
	\begin{equation}
	\label{coord transf retarded}
	X^\mu=u\, n^\mu+r\, \hat q^\mu(x^i)\,,
	\end{equation}
	where $n^\mu$ and $\hat q^\mu(x^i)$ are null vectors with cartesian components given by
	\begin{align}
	\label{hat q}
	\hat q^\mu(x^i)&= \frac{1}{\sqrt{2}}\left(1+x^2\,, 2x^i\,, 1-x^2\right)\,,\\
	n^\mu&=\frac{1}{\sqrt{2}}\left(1,0^i,-1\right)\,,
	\end{align}
	and satisfying $n \cdot \hat q=-1$ as well as $\hat q(x) \cdot \hat q(y)=-|x-y|^2$. In retarded coordinates the flat metric takes the simple form
	\begin{equation}
	\label{flat Bondi gauge}
	ds^2=\eta_{\mu\nu} \dd X^\mu \dd X^\nu =-2 \dd u \dd r+ 2 r^2 \delta_{ij} \dd x^i \dd x^j\,,
	\end{equation}
	such that future null infinity $\scri$ lies at $r \to \infty$. 
	
	As a first step let us recover the Carrollian representation \eqref{Carroll generators} of the Poincar\'e generators from their standard Minkowskian representation. Indeed in Minkowski space $\mathbb{M}^{d+1}$ the Poincaré algebra $\mathfrak{iso}(1,d)$ is naturally represented by the differential operators
	\begin{equation}
	J_{\mu\nu}=i\left(X_\mu\, \frac{\partial}{\partial X^\nu}-X_\nu\, \frac{\partial}{\partial X^\mu}\right)\,, \qquad P_\mu=-i\frac{\partial}{\partial X^\mu} \,,
	\end{equation}
	satisfying
	\begin{equation}
	\begin{split}
	\left[J_{\mu\nu}\,,J_{\rho \sigma}\right]&=-i\left(\eta_{\mu\rho}\, J_{\nu \sigma}+\eta_{\nu \sigma}\,J_{\mu \rho}-\eta_{\mu\sigma}\, J_{\nu\rho}-\eta_{\nu \rho}\, J_{\mu \sigma} \right)\,,\\
	\left[J_{\mu\nu}\,,P_\rho \right]&=-i \left(\eta_{\mu\rho}\, P_\nu-\eta_{\nu\rho}\, P_\mu \right)\,.
	\end{split}
	\end{equation}
	In terms of retarded coordinates, they read
	\begin{equation}
	\begin{split}
	P_0&=-\frac{i}{\sqrt{2}}\left(\partial_r+(1+x^2)\partial_u-r^{-1}x^i\partial_i \right)\,,\\
	P_i&=-\frac{i}{\sqrt{2}}\left(-2x_i\, \partial_u+ r^{-1} \partial_i\right)\,,\\
	P_d&=-\frac{i}{\sqrt{2}}\left(\partial_r-(1-x^2)\partial_u-r^{-1}x^i\partial_i \right)\,,
	\end{split}
	\end{equation}
	and
	\begin{equation}
	\label{J differential}
	\begin{split}
	J_{0d}&=-i (r\partial_r -u\partial_u-x^i \partial_i )\,,\\
	J_{i0}&=\frac{i}{2}\left((r^{-1}u+1+x^2)\partial_i+2 x_i(r\partial_r-u\partial_u-x^j\partial_j) \right)\,,\\
	J_{id}&=\frac{i}{2}\left((r^{-1}u-1+x^2)\partial_i+2 x_i(r\partial_r-u\partial_u-x^j\partial_j) \right)\,,\\
	J_{ij}&=i(x_i \partial_j-x_j \partial_i)\,. 
	\end{split}
	\end{equation}
	It is straightforward to check that their limit $r \to \infty$ exactly coincides with the expression given in \eqref{tilde J}-\eqref{tilde P} for the orbital part of $\tilde P_\mu, \tilde J_{\mu\nu}$, up to the $r \partial_r$ terms which survive in that limit as well. This operator $r \partial_r$ is a geometrical bulk realisation of the conformal weight $\Delta$ as can be observed by comparison with \eqref{Carrollian induced rep}. Said differently, we can already anticipate that a massless bulk field $\Phi$ with asymptotic behaviour
	\begin{equation}
	\label{PHI}
	\Phi(r,x^\alpha) = r^\Delta\, \Phi_\Delta(x^\alpha)+...
	\end{equation}
	will induce a conformal field $\Phi_\Delta(x^\alpha)$ of conformal dimension $\Delta$ at $\scri$. We expect this conformal dimension to be determined by the spin of the representation as in \eqref{Delta(s)}. We will come back to this point momentarily.
	
	Let us make this more precise and study the behavior near $\scri$ of the bulk quantum field $\phi_{\mu_1...\mu_s}$ associated with states in the spin-$s$ helicity representation, and satisfying the covariant field equations \eqref{bulk eom phi}. In retarded coordinates the wave equation takes the form 
	\begin{equation}
	\label{Klein Gordon}
	\square \phi_{\mu_1...\mu_s}=-2\partial_u \nabla_r \phi_{\mu_1...\mu_s}+\nabla^i\nabla_i \phi_{\mu_1...\mu_s}=0\,,
	\end{equation}
	while the transversality and traceless constraints imply
	\begin{align}
	\label{trace transversality}
	g^{ij} \phi_{ij...}=2 \phi_{ur...}\,, \qquad g^{ij} \nabla_i \phi_{j...}=\partial_u \phi_{r...}+\nabla_r \phi_{u...}\,.
	\end{align}
	From this we show that the wave operator takes the form
	\begin{equation}
	\label{Klein Gordon 2}
	\begin{split}
	\square \phi_{r(m) u(s-m-k) i(k)}&=\left(-2\partial_u \partial_r+(2k+1-d-2m)r^{-1} \partial_u+\frac{1}{2r^2}\partial^2 \right)\phi_{r(m) u(s-m-k) i(k)}\\
	&-m \left(2r^{-1}\partial_r+(2m+d-3-2k)r^{-2} \right)\phi_{r(m-1) u(s-m-k+1) i(k)}\\
	&-2k r^{-1} \partial_{(i(1)} \phi_{i(k-1))r(m) u(s-m-k+1) }\\
	&+2k(k-1) \delta_{(i(2)} \phi_{i(k-2))r(m) u(s-m-k+2)}\,,
	\end{split}
	\end{equation}
	where we have introduced the following shorthand notation for the various field components,
	\begin{equation}
	\phi_{r(m)u(s-m-k)i(k)}\equiv \phi_{r_1...r_m\, u_1\,...\,u_{s-m-k}\,i_1\,...\,i_k}\,.
	\end{equation}
	Equation \eqref{Klein Gordon 2} implies that the field component behave asymptotically like
	\begin{equation}
	\phi_{r(m) u(s-m-k) i(k)}(r,x^\alpha)=r^{k-m-\frac{d-1}{2}} \bar \phi_{r(m) u(s-m-k) i(k)}(x^\alpha)+...\,.
	\end{equation}
	The components $\bar \phi_{u(s-k) i(k)}$ are completely unconstrained, while the radial components are determined through the trace and transversality constraints \eqref{trace transversality}.
	Up to this point we have not used the gauge redundancy \eqref{gauge symmetry} which should precisely further reduce the independent physical asymptotic components $\bar \phi_{\mu_1...\mu_s}$ to the tensor components $\bar \phi_{i_1...i_s}$ carrying the spin-$s$ representation of the (short) little group. Indeed the gauge parameter $\varepsilon_{\mu_1...\mu_{s-1}}$ satisfies \eqref{bulk eom xi} and thus similarly behaves asymptotically like
	\begin{equation}
	\varepsilon_{r(m)u(s-k-m-1) i(k)}(r,x^\alpha)=r^{k-m-\frac{d-1}{2}}\, \bar \varepsilon_{r(m)u(s-m-k-1) i(k)}(x^\alpha)+...\,.
	\end{equation}
	Since the components of $\bar \varepsilon_{u(s-k-1)i(k)}$ are completely unconstrained, they allow to gauge away all retarded time components $\bar \phi_{u(s-k)i(k)}$ ($k\neq s$) since the latter transform as
	\begin{equation}
	\delta_{\varepsilon} \bar \phi_{u(s-k)i(k)}=\partial_{u} \bar \varepsilon_{u(s-k-1)i(k)}\,.
	\end{equation}
	Thus the independent gauge-invariant components are just the spatial components $\bar \phi_{i_1...i_s}$. 
	
	We now have a look at the transformation of the gauge-invariant field components $\bar \phi_{i_1...i_s}$ under Poincar\'e transformations. We know that the bulk field $\phi_{\mu_1...\mu_s}$ transforms covariantly, such that it can be expressed in arbitrary coordinates in terms of the Lie derivative
	\begin{equation}
	\label{Lie derivative}
	\delta_\xi \phi_{\mu_1...\mu_s}=\lied_\xi \phi_{\mu_1...\mu_s}=\xi^\rho \partial_\rho \phi_{\mu_1...\mu_s}+\partial_{\mu_1} \xi^\rho\, \phi_{\rho...\mu_s}+...+\partial_{\mu_s} \xi^\rho\, \phi_{\mu_1...\rho}\,,
	\end{equation}
	where $\xi^\mu$ is a Killing vector field. Adopting the standard parametrisation 
	\begin{equation}
	\xi^\mu=A^\mu+\Omega^\mu{}_\nu\, X^\nu\,,
	\end{equation}
	with $A^\mu$ and $\Omega_{\mu\nu}=-\Omega_{\nu\mu}$ corresponding to translations and Lorentz rotations, respectively, its components in retarded coordinates are explicitly given by
	\begin{equation}
	\begin{split}
	\xi^r&=-n^\mu \xi_\mu=-r\, n^\mu \Omega_{\mu\nu} \hat q^\nu-n^\mu A_\mu\,,\\
	\xi^u&=-\hat q^\mu \xi_\mu=-\hat q^\mu A_\mu+u\, n^\mu \Omega_{\mu\nu} \hat q^\nu\,,\\
	\xi^i&=\frac{1}{2r}\partial_i \hat q^\mu\, \xi_\mu=-\frac{1}{2}\left(\hat q^\mu \Omega_{\mu\nu} \partial_i \hat q^\nu+r^{-1}\partial_i \xi^u \right)\,.
	\end{split}
	\end{equation}
	The various Lorentz contractions can be evaluated in terms of the constant cartesian components of $A^\mu$ and $\Omega_{\mu\nu}$, 
	\begin{equation}
	\begin{split}
	\hat q^\mu A_\mu&=-\frac{1}{\sqrt{2}}\left( A^0-A^d+(A^d+A^0)x^2-2 A_i x^i\right)\,,\\
	n^\mu \Omega_{\mu\nu} \hat q^\nu&=\Omega_{0d}+(\Omega_{0i}-\Omega_{di})x^i\,,\\
	\hat q^\mu \Omega_{\mu\nu} \partial_i \hat q^\nu&=-2\Omega_{0d} x_i+(\Omega_{0i}+\Omega_{di})+(\Omega_{0i}-\Omega_{di})x^2-2(\Omega_{0j}-\Omega_{dj})x^j x_i-2\Omega_{ij} x^j\,.
	\end{split}
	\end{equation}
	In the limit $r \to \infty$ we recover the Carrollian vector field \eqref{zetas}
	\begin{equation}
	\begin{split}
	\zeta^u &\equiv \lim_{r \to \infty} \xi^u=a+k x^2+b_i x^i+u(\lambda-2k_i x^i)\,,\\
	\zeta^i&\equiv \lim_{r \to \infty} \zeta^i=a^i+\omega^i{}_j\, x^j+\lambda x^i+k^i x^2-2k_j\, x^j x^i\,,
	\end{split}
	\end{equation}
	upon identifying the Carrollian symmetry parameters
	\begin{align}
	a&=-\frac{1}{\sqrt{2}}(A_0+A_d)\,, &  a_i&=-\frac{1}{2}(\Omega_{0i}+\Omega_{di}) \,, & \lambda&=\Omega_{0d}\,, & b_i&=-\sqrt{2} A_i\,,\\
	\nonumber
	k&=-\frac{1}{2}(A_0-A_d)\,, & k_i&=-\frac{1}{2}(\Omega_{0i}-\Omega_{di})\,, & \omega_{ij}&=\Omega_{ij}\,.
	\end{align}
	In addition we find that the asymptotic limit of the $r$-component is related to the scaling factor \eqref{Omega},
	\begin{equation}
	\xi^r=-r(\lambda-2 k_i x^i)+O(r^0)=-r\, \Omega+O(r^0)\,.
	\end{equation}
	The extra radial direction, or holographic direction, gives a geometrical encoding of the dilation operator $D$. Putting these equations together, we obtain the transformation of the gauge-invariant tensor components, which can be simply written
	\begin{equation}
	\label{Lie phi bar}
	\delta \bar \phi_{i_1...i_s}=\left[\lied_\zeta-\left(s-\frac{d-1}{2}\right)\Omega\right]\bar \phi_{i_1...i_s}\,.
	\end{equation}
	This is nothing but the transformation \eqref{Carrollian conformal transf} of a Carrollian conformal primary field of conformal dimension\footnote{For $d=3$ this result agrees with the findings in \cite{Donnay:2022aba,Donnay:2022wvx}. In that case we have $\Delta=s-1$ and the Carrollian weights $(k,\bar k)$ are determined in terms of the helicity $J=\pm s$ through
		\begin{equation*}
		k+\bar k=s -\Delta=1\,, \qquad k-\bar k=J\,.
		\end{equation*}}
	\begin{equation}
	\Delta(s)=s-\frac{d-1}{2}\,.
	\end{equation}
	Hence \textit{relativistic massless fields on $\mathbb{M}^{d+1}$ are dual to Carrollian conformal primary fields at~$\scri$}. Note also that the combination $-\Delta\Omega$ in \eqref{Lie phi bar} comes from the radial derivative $\xi^r \partial_r=-\Omega\, r\partial_r$ appearing in the Lie derivative \eqref{Lie derivative}, in agreement with the geometrisation $\Delta \mapsto r \partial_r$ described around \eqref{PHI}. 
	
	Using the above correspondence, it is easy to express the Carrollian conformal fields in terms of the creation operators $a_{p,\sigma}^\dagger$. We simply need to consider the expression \eqref{intertwining} in the limit $r \to \infty$. For that we note that the intertwiner wavefunctions are of the form \cite{Weinberg:1995mt}
	\begin{equation}
	\phi^{p,\sigma}_{\mu_1...\mu_s}(X) = \epsilon^{p,\sigma}_{\mu_1...\mu_s}\, e^{-ip \cdot X}\,,
	\end{equation}
	where $e^{-ip\cdot X}$ carries the spacetime dependence, and $\epsilon^{p,\sigma}_{\mu_1...\mu_s}$ is a polarisation tensor with constant cartesian components. If we adopt the convenient parametrisation of a generic null momentum
	\begin{equation}
	\label{momentum parametrisation}
	p^\mu(\omega,y^i)=\omega \hat q^\mu(y^i)\,,
	\end{equation}
	with $\hat q$ given in \eqref{hat q},
	such that the Lorentz-invariant measure becomes 
	\begin{equation}
	d^d p=  \omega^{d-2} \dd \omega \dd^{d-1} y\,,
	\end{equation}
	then the phase of the plane wave takes the simple form
	\begin{equation}
	p(\omega,y) \cdot X(r,u,x)=-\omega(u+r |x-y|^2)\,.
	\end{equation}
	In the limit $r \to \infty$, the $d^{d-1}y$ integral in \eqref{intertwining} can be performed by stationary phase approximation, which localises at $y=x$ and yields
	\begin{equation}
	\phi_{\mu_1...\mu_s}=\left(\frac{-i\pi}{r}\right)^{\frac{d-1}{2}} \sum_{\sigma} \int_0^\infty \dd\omega\, \omega^{\frac{d-3}{2}}\, \epsilon_{\mu_1...\mu_s}^{p,\sigma}\, e^{i\omega u}\, a^\dagger_{p,\sigma}\Big|_{p=\omega \hat q(x^i)}+...
	\end{equation}
	The components of interest are then given by
	\begin{equation}
	\phi_{i_1...i_s}=r^s\, \partial_{i_1} \hat q^{\mu_1}\, ...\, \partial_{i_s} \hat q^{\mu_s}\, \phi_{\mu_1...\mu_s}\,, 
	\end{equation}
	which, in the limit $r \to \infty$, 
	provide the nonzero components of the Carrollian conformal field,
	\begin{equation}
	\bar \phi_{i_1...i_s}=\lim_{r \to \infty} r^{\frac{d-1}{2}-s}\, \phi_{i_1...i_s}=(-i\pi)^{\frac{d-1}{2}} \sum_{\sigma} \int_0^\infty \dd\omega\, \omega^{\frac{d-3}{2}}\, \bar \epsilon_{i_1...i_s}^{p,\sigma}\, e^{i\omega u}\, a^\dagger_{p,\sigma}\Big|_{p=\omega \hat q(x^i)}\,,
	\end{equation}
	with the polarisation tensors
	\begin{equation}
	\bar \epsilon_{i_1...i_s}^{p,\sigma}\equiv \partial_{i_1} \hat q^{\mu_1}\, ...\, \partial_{i_s} \hat q^{\mu_s}\, \epsilon_{\mu_1...\mu_s}^{p,\sigma}\,.
	\end{equation}
	By contrast to massless bulk fields, the dual Carrollian fields are not constrained by a wave equation or analogue. The group theoretical reason for this is that the massless Casimir equation $\mathcal{C}_2=0$ is automatically satisfied in the Carrollian representation of the $ISO(1,d)$ algebra, as shown in \eqref{C2=0}. 
	
	\section{The particles states at null infinity}
	\label{sec:particle states}
	In this section we will show how the particle states  appear on  the boundary $\scri^+$ of Minkowski spacetime. More precisely we will consider the particles as the irreducible representation of the Poincar\'e group, as formulated by Wigner in 1939 \cite{Wigner:1939cj}, and push them to the boundary by taking the limit $r\to \infty$. We will find that the physical   states in momentum space are naturally encoded in   $\scri^+$ even though   this  is  part of spacetime. 
	\par
	To construct an irreducible representation of the Poincare group,  we choose a reference momentum for a massless particle and in particular  $p^{(0)}_\mu=(-1/\sqrt{2}, 0,...,0,1/\sqrt{2})$, or in light-cone notation 
	$p^{(0)}{}^{+}\equiv  p^{(0)}_{-}=1 $ with all other components being zero. We use the notation $V^{\pm}\equiv {1\over \sqrt {2}} (V^{d}\pm V^{0})=V_{\mp}$. The little algebra $\tilde {\cal H}$ that preserves this choice has the generators $\tilde {\cal H}=\{\tilde J_{ij} ,\ \tilde J_{i+}, \ \tilde P_\mu\}$. We begin by taking an irreducible unitary  representation of the little group $\tilde {\cal H}$ which acts on the states  $\psi _\sigma(p^{(0)} )\equiv |p^{(0)}, \sigma\rangle$ with the chosen momentum. As such we take
	\begin{equation}
	\label{5.1}
	\tilde P_{-}\, \psi_\sigma(p^{(0)} )=\psi_\sigma (p^{(0)} )\,, \qquad \tilde P_{+}\, \psi_\sigma (p^{(0)} )=0=\tilde P_{i}\, \psi_\sigma (p^{(0)})\,,
	\end{equation}
	and
	\begin{equation}
	\label{5.2}
	i\tilde J_{ij}\,  \psi_\sigma (p^{(0)})=-\sum_{\sigma^\prime} (D_{ij})_{\sigma}{}^{ \sigma^\prime} \psi_{\sigma^\prime} (p^{(0)})\,, \qquad \tilde J_{i+}\, \psi_\sigma (p^{(0)} ) =0\,.
	\end{equation} 
	The last equation is required by unitarity of the representation as the $\tilde J_{i+}$  form an Abelian algebra. Note that the 
	translation generators $\tilde P_\mu$ should rightly be thought of as part of the little group as the states $\psi_\sigma (p^{(0)} )$ also carry a representation of these. In what follows we will take this to be understood and use the passive action for the generators. The discussion of the irreducible representation of the Poincar\'e group in this section has considerably benefitted from reference \cite{PeterPeter}. 
	\par
	The states in the full representation are found by boosting the above states by the action of the generators of the Lorentz algebra which are not in the little algebra, that is, the generators $\tilde J_{-i}$ and $\tilde J_{+-}$. We define 
	\begin{equation}
	\label{5.3}
	\psi_\sigma (p) \equiv  e^{i\varphi^i \tilde J_{-i}}\, e^{i\phi \tilde J_{+-}}\, \psi _\sigma(p^{(0)})\,.
	\end{equation}
	To determine the momentum $p_\mu$  of this state we just act with $\tilde P_\mu$, namely 
	\begin{equation}
	\tilde P_\mu\, \psi _\sigma (p)=  (e^{i\varphi^i \tilde J_{-i}} e^{i\phi \tilde J_{+-}} e^{-i\phi \tilde J_{+-}} e^{-i \varphi^i \tilde J_{-i}} \tilde P_\mu\, e^{i\varphi^i \tilde J_{-i}} e^{i\phi \tilde J_{+-}})\, \psi_\sigma (p^{(0)})
	= p_\mu\,  \psi _\sigma (p)\,.
	\end{equation}
	Using the commutation relations of the Poincar\'e group and equation \eqref{5.1} we find
	\begin{equation}
	\label{5.5}
	p^+= e^\phi\,, \qquad  p^-= -{1\over 2} \varphi^i\varphi_i e^\phi \,, \qquad p^i= -\varphi^i e^\phi\,.
	\end{equation}
	Thus we find how the $d+1-1=d$ components of the massless momenta are parametrised by the $d$ group parameters $\phi$ and $\varphi^i$. Identifying $e^\phi=\omega$ and $\varphi^i=-\sqrt {2}y^i$ we recognise precisely the parametrisation of equation \eqref{momentum parametrisation}. 
	It is interesting to see how the parametrisation of the momenta, that is,  $(\omega, y^i)$  associated with  $\scri^+$ arises naturally in  the construction of the irreducible representations of the Poincar\'e group. 
	\par
	The states in the irreducible representation transform under the Lorentz group, $g \in SO(1,d)$, as 
	$$ 
	U(g) \psi_\sigma (p) = D (h^{-1} ) _\sigma {}^{\sigma^\prime} \psi_{\sigma^\prime} (p^\prime) 
	\eqno(5.6)$$
	where $h\in \tilde {\cal H}$ is defined by the coset relation
	\begin{equation}
	g\, e^{i\varphi^i \tilde J_{-i}} e^{i\phi \tilde J_{+-}}= e^{i\varphi^i{}^\prime \tilde J_{-i}} e^{i\phi^\prime \tilde J_{+-}}\, h\,,
	\end{equation}
	and $p^\mu {}^\prime$ is the momentum corresponding to $\varphi^i{}^\prime$ and $\phi^\prime$. 
	Indeed if we take our group element $g$ to be of the form 
	\begin{equation}
	\label{5.7}
	g= e^{-{i\over 2} \Omega ^{\mu\nu} \tilde J_{\mu\nu}}= e^{-i\lambda \tilde J_{+-} -i\sqrt {2}a^i \tilde J_{-i}-{i\over 2}\omega^{ij} \tilde J_{ij}+i\sqrt {2}k^i \tilde J_{+i}}\,,
	\end{equation}
	we find that 
	\begin{equation}
	\begin{split}
	\varphi^i{}^\prime&= \varphi^i -\sqrt {2}a^i+\omega^i{}_j\varphi^j  -\lambda \varphi^i +\sqrt {2}(\varphi^j k_j)  \varphi^i -{1\over \sqrt {2}} ( \varphi^j  \varphi_j ) k^i\,,\\
	\phi^\prime &=\phi +\lambda -\sqrt {2}(k^i \varphi_i)\,,
	\end{split}
	\end{equation}
	with
	\begin{equation}
	\label{5.9}
	h=e^{-{i\over 2}\omega^{ij} \tilde J_{ij} -i\sqrt {2}k^i\varphi^j \tilde J_{ij} + ie^{-\phi } \sqrt {2}k^i \tilde J_{+i}}\,.
	\end{equation}
	\par
	We can also compute the action of the translation $g=e^{-ia\tilde P_- -ik\tilde P_+ - {i\over \sqrt {2}} b^i \tilde P_i}$ on $\psi_\sigma (p)$ by passing it though the factor $e^{i\varphi^i \tilde J_{-i}} e^{i\phi \tilde J_{+-}}$ and using equation \eqref{5.1}, one finds that 
	\begin{equation}
	\label{5.10}
	e^{-ia\tilde P_- -ik\tilde P_+-{i\over \sqrt {2}} b^i \tilde P_i}\psi (p) = e^{-i\omega (a + {k\over 2} \varphi^j\varphi _j -{1\over \sqrt {2}}\varphi^j b_j)} \psi (p) 
	\end{equation}
	To find this result we used the identity
	\begin{equation}
	e^{A+B}= e^B e^{A+{1\over 2} [A, B]-{1\over 12} [A, [ A, B]]+\ldots } = e^{A-{1\over 2} [A, B]-{1\over 12} [A, [ A, B]]+\ldots }e^B\,,
	\end{equation}
	valid for any  for two generators $A$ and $B$ but for only first order in $B$. In this particularly simple case one can also use equation \eqref{X prime} rather than the above more complicated identity.
	\par
	In order to push these particle states to the boundary we require them in the Minkowski spacetime and so we take the Fourier transform
	\begin{equation}
	\psi_\sigma (X) \equiv \int {d^{d+1} p\over {(2\pi )}^{d+1}}  2\pi \delta (p^2)\theta (p^0)   e^{ip\cdot X} \psi_\sigma (p)
	= \int {d^d p\over 
		2 p^0 {(2\pi)}^d} e^{ip\cdot X} \psi_\sigma (p)\,.
	\end{equation}
	Using relation between the coordinates  $X^\mu$ of Minkowski space $\mathbb{M}^{d+1}$ and the coordinates $(r,u,x^i)$  of equation \eqref{coord transf retarded}, which are suited to the emergence of $\scri^+$, we find that 
	\begin{equation}
	e^{ip\cdot X} = e^{-i\omega u} e^{-ir\omega (x-y)^2}\,,
	\end{equation}
	while the change of variable from $p^i$ to $\omega, y^i$ has Jacobian ${d^d p \over p^0}= (\sqrt {2} )^{d-1} \omega^{d-2} d\omega\, dy^i$. Using these results we find that 
	\begin{equation}
	\psi_\sigma (X) = \int {dy^i \over (2\pi)^d}{ d\omega\over \omega}  (\sqrt {2})^{d-3} \omega^{d-1}  e^{-i\omega u} e^{-ir\omega (x-y)^2} 
	e^{i\varphi^i \tilde J_{-i}}e^{i\phi \tilde J_{+-}}\, \psi_\sigma (p^{(0)})\,.
	\end{equation}
	We will now take the limit of $r \to \infty$ of $\psi_\sigma (X)$ to obtain the states at the boundary $\scri^+$ by using the formula
	\begin{align}
	\lim_{r\to \infty} \int _a^b dx f(x) e^{irg(x) }= \lim_{r\to \infty} f(z) e^{irg(z)} \sqrt{{2\pi i \over r  g^\prime{}^{\prime} (z)}}\,,
	\end{align}
	where $z$ is a point in the interval $[a,b]$ where $g^\prime (z)=0$ and $g^\prime{}^{\prime} (x)= {d^2 g(x) \over dx^2}$. 
	We find that 
	\begin{equation}
	\psi_\sigma (u, x^i )\equiv \lim_{r\to \infty} r^{{d-1}\over 2}\psi_\sigma (X)= c_1 \int _0^\infty d\omega\, \omega^{{d-3\over 2}} e^{-i\omega u} e^{ix^i P_i} e^{-i\ln \omega D}\, \psi_\sigma (p^{(0)})\,,
	\end{equation}
	where $c_1= {1\over 2} e^{-i{(d-1)\over 2}\pi } (2\pi)^{-{d+1\over 2}}$ and we have relabelled the generators by $P_i= \sqrt{2}\, \tilde J_{i-}$ and $\tilde J_{+-}=-D$. In the above equation we can write $e^{ix^i P_i} e^{-i\ln \omega D}\, \psi_\sigma (p^{(0)} )= \psi_\sigma (p )= \psi_\sigma (\varphi^i, \phi )= \psi_\sigma ( x^i , \omega )$ where by abuse of notation we use the same symbol $ \psi$ for the function even though we have changed the variables. 
	\par
	To better understand what these states are we consider them at the origin of $\scri^+$, namely at $u=0=x^i$. One readily finds that
	\begin{equation}
	\label{5.17}
	K_i\, \psi_\sigma (0,0)=0=  K\, \psi_\sigma (0,0)=B_i\, \psi_\sigma (0,0)\,, 
	\end{equation}
	where we have relabelled the Poincare generators as follows 
	\begin{equation}
	\label{5.18}
	K_i=-\sqrt{2}\, \tilde J_{i+}\,, \qquad K=\tilde P_+\,, \qquad \sqrt{2}\, B_i=- \tilde P_i\,.
	\end{equation}
	Acting with $D$ we find that 
	\begin{equation}
	\label{5.19}
	iD\, \psi_\sigma(0,0 )= -c_1\int d\phi\, e^{{(d-1)\over 2}\phi}\frac{d}{d\phi} e^{-i\phi D} \psi_\sigma (p^{(0)} )
	= {(d-1)\over 2} \psi_\sigma (0,0)\,.
	\end{equation}
	This is essentially the same as \eqref{action D} together with \eqref{Delta(s)}, however the two expressions differ by the spin $s$ factor. The reason for this difference comes from the $r$-dependence of the relativistic bulk fields and their tensorial transformations to which we will come back at the end of this section.
	We also note that 
	\begin{equation}
	\label{5.20}
	e^{{u\over \sqrt {2}} iH+ x^i iP_i}  \psi_\sigma (0,0)=  \psi_\sigma (u,x^i )\,, \qquad iJ_{ij}\, \psi_\sigma (0,0 ) = -\sum_{\sigma^\prime} (D_{ij})_{\sigma}{}^{ \sigma^\prime}\, \psi_{\sigma^\prime} (0,0)\,,
	\end{equation}
	where $H=-\tilde P_{-}$. To derive the last equation we used the identity $e^{-i\phi \tilde J_{+-} } \tilde P_{-}\, e^{i\phi \tilde  J_{+-} }= e^\phi \tilde P_{-} $. 
	\par
	Let us summarise the situation. We began with an irreducible representation of the Poincar\'e algebra with the generators $\tilde J_{\mu\nu}, \tilde P_\mu$ corresponding to  a massless particle in Minkowski spacetime. We pushed these states to the boundary $\scri^+$  of Minkowski spacetime by taking the $r\to \infty $ on their Fourier transform. We found the states $ \psi_\sigma (u,x^i )$ living on $\scri^+$ which obey equations \eqref{5.17}, \eqref{5.19} and \eqref{5.20} and also carry a representation of the Poincar\'e algebra but with the generators identified as 
	\begin{equation}
	\label{5.21}
	{\cal H}=\{ J_{ij}=\tilde J_{ij} ,\ K_i=-\sqrt {2} \tilde J_{i+} ,\ K=\tilde P_+ , \ \sqrt{2} B_i= -\tilde P_i,  \ D=-\tilde J_{+-}\}\,,
	\end{equation}
	as well as 
	\begin{equation}
	\label{5.22}
	H= -\tilde P_{-}\,, \qquad P_i= -\sqrt {2} \tilde J_{-i}\,.
	\end{equation}
	We recognise  $ \psi_\sigma (u,x^i )$ as a representation  induced  from  $ \psi_\sigma(0,0 )$ which carries a representation of the subalgebra $ {\cal H}$ which is boosted by the generators $H$ and $P_i$. We observe that the subalgebra  $ {\cal H}$ has the extra generator $D$ compared to the little algebra $\tilde  {\cal H}$ for the states in the Wigner construction. The construction involving $ \psi_\sigma (u,x^i )$ is one  typically associated with  a representation of the conformal group  but  in this case this group is  the Poincar\'e group. In this process   the translation generators on $\scri^+$ are some of the Lorentz generators in Minkowski spacetime, see  equations \eqref{5.21} and \eqref{5.22}. 
	\par
	It is instructive to find a general Lorentz transformation of $\psi_\sigma (x^i, u)$, using equation \eqref{5.7} we find that 
	\begin{equation}
	\begin{split}
	U(g) \psi_\sigma (x^i, u)= c_1 \int _0^\infty d\omega\, \omega^{{d-3\over 2}} e^{-i\omega u} e^{i\varphi^{i \prime}\tilde J_{-i} } e^{i\phi^\prime  \tilde J_{+-}} D(h^{-1})_\sigma {}^{\sigma^\prime}  \psi_{\sigma^\prime} (p^{(0)} )\\
	=c_1 \int _0^\infty d\omega^\prime (\omega^\prime)^{{d-3\over 2}} e^{-i\omega^\prime e^{(\lambda +\sqrt {2}\varphi^i k_i)}u}\, 
	e^{{(d-1)\over 2}(\lambda+\sqrt {2}\varphi^ik_i)}D(h^{-1})_\sigma {}^{\sigma^\prime}e^{i\varphi^{i \prime}\tilde J_{-i} } e^{i\phi^\prime  \tilde J_{+-}}\,  \psi_{\sigma^\prime} (p^{(0)} )\,,
	\end{split}
	\end{equation}
	where $\omega^\prime= e^{\phi^\prime}= e^{-(\lambda +\sqrt {2}\varphi^i k_i)}\omega$. Using the above  identification $x^i=y^i= -{\varphi^i\over \sqrt {2}}$ and relabelling the integration over $\omega$ we find that 
	\begin{equation}
	U(g) \psi_\sigma (x^i, u)=  \psi_\sigma (x^{i\prime},  u^\prime)+{(d-1)\over 2}(\lambda - {2}x^i k_i) \psi_\sigma (x^i, u)
	+ ({\omega^{ij}\over 2} -{2} k^ix^j )D(J_{ij})_\sigma {}^{\sigma^\prime}  \psi_{\sigma^\prime} (x^i, u )\,,
	\end{equation}
	where $x^{i\prime}$ and $u^\prime$ are given by 
	\begin{equation}
	\label{5.25}
	\begin{split}
	x^i{}^\prime&= x^i +a^i+\omega^i {}_j x^j +\lambda x^i - {2}(x^j k_j)  x^i + ( x^j  x_j ) k^i\,,\\
	u^\prime &=u+\lambda u -{2} (k^i x_i) u +a + k x^jx_j +b_j x^j\,,
	\end{split}
	\end{equation}
	where we have used equation \eqref{5.10} to find  the last transformations of $u$. These results agree with those of equation \eqref{zetas}. 
	\par
	The boost is parametrised by the $d$ variables $\varphi^i$ and $\phi$ while  $\scri^+$ has the $d$ coordinates $(x^i , u)$. The $\varphi^i$  and the coordinates $x^i $ are related by  $\varphi ^i= -\sqrt {2} x^i$ while  $u$ and  $\omega=e^\phi$ are conjugate variables in the    Fourier transform. Hence even though the $\varphi^i$ arise in momentum space and the $x^i$ parametrise  the asymptotic region of spacetime they are the same up to a multiplicative constant. Indeed $\scri^+$ encodes the kinematics of the particle in a way that is not subject to any constraint and it carries in essence the same irreducible representation of the Poincar\'e group. 
	\par
	In relativistic quantum field theory we usually prefer to work with quantities that transform in a way which makes their Poincar\'e symmetry manifest. 
	To do this we consider  a finite dimensional, albeit non-unitary,  representation of the Lorentz group that contains the above irreducible representation when we restrict to the little group $\tilde {\cal H}$. Let us  denote these fields by $\Psi_n$ and the matrix of the finite dimensional non-unitary representation by $\tilde D(g)_n{}^m$. However, we only use this representation for transformations of the little group $\tilde {\cal H}$  and we construct the induced representation in the same way as above.  For  the chosen momenta $ p^{(0)}_{-}=1 $  we demand that under the little group $\tilde {\cal H}$  it transforms as 
	\begin{equation}
	\label{5.26}
	i\tilde J_{ij}  \Psi_n (p^{(0)})=-\sum_{m} (\tilde D_{ij})_{n}{}^{ m} \Psi_m (p^{(0)} )\,, \quad i\tilde J_{+i}\Psi_n (p^{(0)} ) =-\sum_{m} (\tilde D_{+i})_{n}{}^{ m} \Psi_m (p^{(0)} )\,,
	\end{equation}
	and we define the states with any momentum by  the same boost as before
	\begin{equation}
	\label{5.27}
	\Psi_n (p ) \equiv  e^{i\varphi^i \tilde J_{-i}} e^{i\phi \tilde J_{+-}}\, \Psi_n (p^{(0)} )\,.
	\end{equation}
	The momentum $p^\mu$ is given by equation \eqref{5.5}. 
	\par
	One then finds that under a general group element $g\in SO(1,3)$ the states transform as 
	\begin{equation}
	U(g) \Psi_n(p) = \sum_m \tilde D (h^{-1} ) _n {}^{m} \Psi_m (p^\prime)\,, 
	\end{equation}
	where $h$ is given in equation \eqref{5.9}. Since $h$ only involves the generators $\tilde J_{ij}$ and $ \tilde J_{+i}$,  which involve  Lorentz transformations with parameters $\Lambda_{ij}$ and  $\Lambda_{-j}$, the component $ \Psi_+$ does not transform,  illustrating the fact that the $\Psi_n$  do not form  an irreducible representation. 
	\par
	To obtain the corresponding fields on $\scri^+$ we have to take the Fourier transform and take the limit $r\to \infty$, as above, to find that 
	\begin{equation}
	\Psi_n ( x^i , u)\equiv \lim_{r\to \infty} c_1  \int _0^\infty d\omega\, \omega^{{d-3\over 2}} e^{-i\omega u}\, \Psi_\mu (p)
	\end{equation}
	where $\Psi_\mu (p)$ is given in equation \eqref{5.25} and can be written after the $r\to \infty$ as $\Psi_\mu (p)=e^{ix^iP_i} e^{-i \ln \omega D}\, \Psi_n ( p^{(0)}) $. Carrying out a general Lorentz transformation in the same way as we did for the physical states alone we find that 
	\begin{equation}
	\label{5.30}
	\begin{split}
	U(g) \Psi_n (x^i, u)=  \psi_n (x^{i\prime},  u^\prime)+{(d-1)\over 2}(\lambda - {2}x^i k_i) \Psi_n (x^i, u)\\
	+ ({\omega^{ij}\over 2} -{2} k^ix^j )\tilde D(J_{ij})_n {}^{m}  \Psi_{m} (x^i, u )
	+{i\sqrt {2}k^i\over \partial_u}\tilde D(J_{+i})_n {}^{m}  \Psi_{m} (x^i, u )
	\end{split}
	\end{equation}
	where $x^{i\prime}$ and $ u^\prime$ are as in equation \eqref{5.25}  and the expression ${1\over i\partial_u}$ just leads to the factor ${1\over \omega}$  in momentum space that occurs in the group element $h$. As we will see this last term does not occur in the transformation of the physical states. 
	\par
	Clearly these fields do not transform in a way that makes the symmetry manifest but we can construct some that do, specifically 
	\begin{equation}
	\label{5.31}
	A_n(p)\equiv \tilde D(e^{i\varphi^i \tilde J_{-i}} e^{i\phi \tilde J_{+-}})_n{}^m\, \Psi_m (p)\,.
	\end{equation}
	Indeed we find that they transform as $U(g) A_n(p) = \tilde D(g^{-1})_n{}^m A_m(p^\prime)$ under a Lorentz transformation. Taking the Fourier transform of $A_n(p)$ and the $r\to \infty $ limit we find  $A_n(x^i, u)$ on  $\scri^+$, namely
	\begin{equation}
	\label{5.32}
	A_n(x^i, u) \equiv \lim_{r\to \infty} c_1  \int _0^\infty d\omega\, \omega^{{d-3\over 2}} 
	e^{-i\omega u} A_n (p)\,, 
	\end{equation}
	where $A_n(p)$ is given by equation \eqref{5.31}. Carrying out a Lorentz transformation we find that $U(g) A_n(x^i, u)= \tilde D(g^{-1})_n{}^m A_m(x^{i\prime} , u^\prime)$.  
	\par
	The procedure is best  illustrated by  an example and we choose that for a spin one particle. The corresponding irreducible representation has  the $d-1$  states $\psi_i (p^{(0)})$ which transform under $SO(d-1)$, obey equations \eqref{5.1} and \eqref{5.2}, and are boosted as in equation \eqref{5.3}. The simplest embedding to find a covariant description  is  in a vector representation which we denote by   $\Psi_\mu$. Using equation \eqref{5.26}  we find that in momentum space it   transforms  under the Lorentz transformation as 
	\begin{equation}
	\label{5.33}
	\begin{split}
	U(g)  \Psi_i(p)=  \Psi_i(p^\prime )-\omega_i{}^j \Psi_j (p) -\sqrt {2}(k^i\varphi^j-k^i\varphi^j) \Psi_j (p)-\sqrt {2}k^ie^{-\phi} \Psi_+ (p)\,,\\
	U(g) \Psi_-(p)=  \Psi_-(p^\prime )-e^{-\phi}\sqrt {2} k^j\Psi_j (p)\,, \quad U(g) \Psi_+(p)= \Psi_+(p^\prime )\,.
	\end{split}
	\end{equation}
	where $p^\mu$ is parametrised by $\varphi^i$ and $\phi$ as in equation \eqref{5.5} and their transformation is given in equation \eqref{5.9}. The transformation of $\Psi_\mu (x^i , u)$ on $\scri^+$ is easily read off from equation \eqref{5.30}. 
	\par
	The corresponding covariant field is denoted by $A_\mu$ which is the familiar Maxwell field. It is related to $\Psi_\mu$ by equation \eqref{5.31} which in this case is given by 
	\begin{equation}
	\label{5.34}
	\begin{split}
	A_i (p)&= \Psi_i(p)-\varphi_i e^{\phi}  \Psi_-(p)\,, \qquad A_-(p)=e^{\phi} \Psi_-(p)\,,\\ 
	A_+(p)&= e^{-\phi} \Psi_+ (p)+\varphi^i \Psi_i (p)-{1\over 2}\varphi^i\varphi_i e^{\phi} \Psi_-(p)\,.
	\end{split}
	\end{equation}
	The inverse transform is given by 
	\begin{equation}
	\label{5.35}
	\begin{split}
	\Psi_i (p)&=\tilde A_i (p)+\varphi_i A_-(p)\,, \qquad  \Psi_-(p)=e^{-\phi} A_-(p)\,,\\ 
	\Psi_+(p)&=e^{\phi} ( A_+ (p)-\varphi^i A_i (p)-{1\over 2}\varphi^i\varphi_i e^{\phi} A_-(p))\,.
	\end{split}
	\end{equation}
	Using equations \eqref{5.33} and \eqref{5.34} one can easily verify that $\delta A_\mu = \Lambda_\mu{}^\nu A_\nu $.  Using equation \eqref{5.34} in the equation \eqref{5.32} for $A_n(x^i, u)$  for we find that 
	\begin{equation}
	\label{5.36}
	\begin{split}
	A_i (x^i , u) =\Psi_i (x^i , u)+i\sqrt{2}x_i\partial_u \Psi_- (x^i , u) ,\ \ A_- (x^i , u)= i\partial_u \Psi_- (x^i , u)\,,\\
	A_+(x^i , u)= {1\over i\partial_u} \Psi_+(x^i , u)-\sqrt{2} x^i\Psi_i (x^i , u) -x^i x_i i\partial_u \Psi_- (x^i,u)\,.
	\end{split}
	\end{equation}
	\par
	Having embedded the irreducible representation into a larger representation we must implement conditions that ensure that it really only contains the original  $d-1$  states. For the case of spin one we should impose that  $p^\mu A_\mu (p) =0$. Using equations \eqref{5.5} and \eqref{5.34} we find this  implies that $\Psi_+(p)=0$,  the other fields being unaffected. Thus  we are left with  the  $d$ fields $\Psi_i (p) $ and $\Psi_-(p) $ which transform into each others under the general  Lorentz transformation of equation \eqref{5.33}. We also require the gauge symmetry $\delta A_\mu (p) = p_\mu \Lambda(p)$ and using equations \eqref{5.5} and \eqref{5.35} we find that it implies that $\delta \Psi_- (p) =\Lambda$, $\delta \Psi_i (p) =0$ and $\delta \Psi_+(p)  =0$. We can use this gauge transformation to set $\delta \Psi_- (p) =0$ leaving us with the $d-1$ physical states. Carrying out the Fourier transform to Minkowski spacetime we find that  $\Psi_+(p)=0$ implies  that  $\Psi_+(x^i , u )=0$ and taking the $r\to \infty$ limit we find that on  $\scri^+$ we have the $d$  unconstrained fields $\Psi_i ( x^i , u)$ and $\Psi_- ( x^i , u)$  which transform into each other under Lorentz transformations and also inherit the gauge symmetry. 
	\par
	We will now make the connection with the fields studied in section~\ref{sec:bulk} where the  usual coordinates $X^\mu$ of Minkowski spacetime were exchanged for the coordinates $x^a= (r,x^i,u)$ and taking the $r\to \infty $ the boundary $\scri^+$ emerged with the coordinates $x^i,u$. The gauge fields $A_\mu^{\scri}$ on $\scri^+$ are then  found from those on Minkowski spacetime by the change of coordinates $A_\mu = {\partial x^a \over \partial X^\mu}A_a^{\scri}$. One finds that 
	\begin{equation}
	\begin{split}
	A_i (x^i , u)& = {1\over \sqrt{2} r} A_i^{\scri}(x^i , u) -\sqrt{2} x^i A_u^{\scri} (x^i , u)\,, \qquad 
	A_-(x^i , u) = -A_u^{\scri} (x^i , u)\,,\\
	A_+(x^i , u)&=  A_r^{\scri}(x^i , u) -{1\over r} x^i A_i^{\scri} (x^i , u) +x^i x_i A_u^{\scri}(x^i , u)\,.
	\end{split}
	\end{equation}
	The inverse transformation is given by 
	\begin{equation}
	\label{5.39}
	\begin{split}
	A_i^{\scri}(x^i , u) &= \sqrt {2} r A_i (x^i , u) -2r x^i A_- (x^i , u)\,, \qquad  
	A_u^{\scri}(x^i , u) = -A_- (x^i , u)\,,\\
	A_r^{\scri}(x^i , u)& = A_+(x^i , u)  +\sqrt {2} x^i A_i (x^i , u) - x^ix_i A_-(x^i , u)\,. 
	\end{split}
	\end{equation}
	Using  equation \eqref{5.36} in  equation \eqref{5.39} we find that 
	\begin{equation}
	\label{5.40}
	A_i^{\scri}(x^i , u) = \sqrt {2}r \Psi_i (x^i , u),\ \ 
	A_u^{\scri}(x^i , u) = i\partial_u \Psi_- (x^i , u),\ \ 
	A_r^{\scri}(x^i , u) = {1\over i\partial_u} \Psi _+ (x^i , u)\,.
	\end{equation}
	Since $\Psi _+=0$ we find $A_r^{\scri}(x^i , u) =0$ leaving us with the fields $\Psi_i$ and $\Psi_-$. As the above equation makes clear these are  very closely related to the fields $A_i^{\scri}$ and $A_u^{\scri}$ discussed in section~\ref{sec:boundary} and section~\ref{sec:bulk}. Note that the extra factor of $r$ in $A_i^{\scri}$ compared to $\Psi_i$ explains the difference in the expression for the dilation operator $D$ acting on $ A_i^{\scri}(x^i , u) $ in equation \eqref{Delta(s)} and on  
	$\Psi_i (x^i , u)$ in equation \eqref{5.30}. Indeed the operator $D$ acts non-trivially on $r$ as we discussed around \eqref{PHI}. 
	
	The last term in the transformation  of $\Psi_\mu$ in equation \eqref{5.30} contains a ${1\over i\partial_u}$  and corresponds to a transformation with a  parameter of the generic form $\Lambda_{-i} $.  It transforms $\Psi_i$ into $\Psi_+$ but as $\Psi_+$ vanishes this contribution vanishes. As such it does not affect the physical states $\Psi_i$. It can also  transform $\Psi_-$ into $\Psi_i$. However, the ${1\over i\partial_u}$ factor disappears in the corresponding transformation of $A_u^{\scri}(x^i , u)$ as equation \eqref{5.40} has a $ i\partial_u $ in the relation between the two fields.
	
	Although we have only explicitly carried out the analysis for the spin one particle the general picture is clear. 
	
	\section*{Acknowledgments}
	We thank Dio Anninos for valuable discussions, and Adrien Fiorucci and Romain Ruzziconi for clarifying the relation to their work (see footnote 4). This work was supported by the STFC grants ST/P000258/1 and ST/T000759/1.
	
	\appendix
	
	\section{Poincar\'e as ultra-relativistic conformal algebra}
	\label{app:ultra-relativistic limit}
	We review how the algebra $\mathfrak{iso}(1,d)$ arises as an In\"on\"u-Wigner contraction of the conformal algebra $\mathfrak{so}(2,d)$, viewed as the ultra-relativistic/Carrollian limit $c \to 0$. We implement this in the following way. Starting with the canonical $d$-dimensional Minkowskian metric 
	\begin{equation}
	ds^2_{\mathbb{M}^d}=\eta_{\alpha\beta} \dd x^\alpha \dd x^\beta=-(\dd x^0)^2+\delta_{ij} \dd x^i \dd x^j\,,
	\end{equation}
	we restore the dependence on the speed of light $c$ via $x^0=c\, u$ and subsequently take the limit $c \to 0$,
	\begin{equation}
	ds^2_{\mathbb{M}^d}=-c^2 \dd u^2+\delta_{ij} \dd x^i \dd x^j \quad \to \quad ds^2_{\scri}=q_{\alpha\beta} \dd x^\alpha \dd x^\beta=0\, \dd u^2 +\delta_{ij} \dd x^i \dd x^j\,.
	\end{equation}
	In this way we obtain $\scri$ as an ultra-relativistic limit of $\mathbb{M}^{d}$. Now let's see how the relativistic conformal algebra $\mathfrak{so}(2,d)$ contracts to the Carrollian conformal algebra $\mathfrak{iso}(1,d)$.
	
	\paragraph{Relativistic conformal algebra.} Thus we start by considering $SO(2,d)$ as the group of conformal isometries of Minkowski space $\mathbb{M}^{d}$. These correspond to the infinitesimal coordinate transformations
	\begin{align}
	\label{x'}
	x'^{\alpha}=x^\alpha+\xi^\alpha\,,
	\end{align}
	with
	\begin{equation}
	\label{xi conformal}
	\xi^\alpha=a^\alpha+\omega^\alpha{}_\beta\,  x^\beta+\lambda x^\alpha+ k^\alpha x^2-2 (k \cdot x) x^\alpha\,.
	\end{equation}
	These comprise translations $a_\alpha$, Lorentz rotations $\omega_{\alpha\beta}$, dilation $\lambda$ and special conformal transformations $k_\alpha$. In terms of symmetry generators, the transformation \eqref{x'} can be written
	\begin{align}
	x'^{\gamma}=(1+ia^\alpha P_\alpha+\frac{i}{2} \omega^{\alpha\beta}J_{\alpha\beta}+i\lambda D+ik^\alpha K_\alpha)\, x^\gamma\,,
	\end{align}
	with
	\begin{align}
	\label{Poincare generators}
	P_\alpha=-i\partial_\alpha\,, \quad
	J_{\alpha\beta}=i(x_\alpha \partial_\beta-x_\beta \partial_\alpha)\,, \quad
	D=-ix^\alpha \partial_\alpha\,, \quad
	K_\alpha=i(-x^2 \partial_\alpha+2x_\alpha\, x^\beta \partial_\beta)\,.
	\end{align}
	They satisfy the $\mathfrak{so}(2,d)$ algebra
	\begin{equation}
	\label{SO(2,d) algebra}
	\begin{split}
	\left[J_{\alpha\beta}\,,J_{\gamma \delta}\right]&=-i\left(\eta_{\alpha\gamma}\, J_{\beta \delta}+\eta_{\beta \delta}\,J_{\alpha \gamma}-\eta_{\alpha\delta}\, J_{\beta\gamma}-\eta_{\beta \gamma}\, J_{\alpha \delta} \right)\,,\\
	\left[J_{\alpha\beta}\,,P_\gamma \right]&=-i \left(\eta_{\alpha\gamma}\, P_\beta-\eta_{\beta\gamma}\, P_\alpha \right)\,,\\
	\left[J_{\alpha\beta}\,,K_\gamma \right]&=-i \left(\eta_{\alpha\gamma}\, K_\beta-\eta_{\beta\gamma}\, K_\alpha \right)\,,\\
	\left[D\,, P_\alpha \right]&=i P_\alpha\,,\\
	\left[D\,, K_\alpha \right]&=-i K_\alpha\,,\\
	\left[K_\alpha\,, P_\beta\right]&=-2i\left(\eta_{\alpha\beta}\, D-J_{\alpha\beta} \right)\,.
	\end{split}
	\end{equation}
	
	\paragraph{Ultra-relativistic conformal algebra.} Now we study the ultra-relativistic limit of the conformal isometries \eqref{x'}. The vector field can be split into time and space components,
	\begin{equation}
	\begin{split}
	\xi^0&=a^0+\omega^0{}_i\, x^i+\lambda x^0+k^0(-(x^0)^2+\vec{x} \cdot \vec x)-2(-k^0 x^0+ \vec k \cdot \vec x)\, x^0\,,\\
	\xi^i&=a^i+\omega^i{}_0\, x^0+\omega^i{}_j\, x^j+\lambda x^i+k^i(-(x^0)^2+\vec{x} \cdot \vec x)-2(-k^0 x^0+ \vec k \cdot \vec x)\, x^i\,.
	\end{split}
	\end{equation}
	Performing the rescaling 
	\begin{align}
	x^0=c\, u\,, \qquad a^0=c\, a\,, \qquad \omega^{0}{}_i=c\, b_i\,, \qquad k^0=c\, k\,,
	\end{align}
	in the limit $c \to 0$ we obtain
	\begin{equation}
	u'=u+\zeta^u\,, \qquad x'^i=x^i+\zeta^i\,,
	\end{equation}
	with 
	\begin{equation}
	\begin{split}
	\zeta^u&=\lim_{c\to 0} c^{-1} \xi^0=a+b_i x^i+k x^2+(\lambda-2 k_i x^i)u\,,\\
	\zeta^i&=\lim_{c \to 0} \xi^i=a^i+\omega^i{}_j\, x^j+\lambda x^i+k^i x^2-2k_j x^j x^i\,.
	\end{split}
	\end{equation}
	The contraction of the Poincar\'e transformations was discussed in the seminal work of L\'evy--Leblond \cite{Levy1965}, which we simply extended to conformal transformations. The ultra-relativistic generators \eqref{Carroll generators} and algebra \eqref{conformal Carroll algebra} presented in section~\ref{sec:boundary} can also be directly obtained by In\"on\"u-Wigner contraction of the relativistic conformal algebra \eqref{SO(2,d) algebra}. In practice we simply have to perform the change of variable $x^0=c\, u$ in the expression of the $\mathfrak{so}(2,d)$ generators and rescale some of them appropriately, namely
	\begin{align}
	H=c P_0\,, \qquad B_i=c J_{0i}\,, \qquad K=c K_0\,,
	\end{align}
	such that the limit $c \to 0$ can be taken and the expressions \eqref{Carroll generators}-\eqref{conformal Carroll algebra} are recovered.
	
	\bibliography{bibl}
	\bibliographystyle{JHEP}  
\end{document}